\documentstyle[12pt,epsf]{article}

  \def\pp{{\mathchoice
          {
              \kern 1pt%
              \raise 1pt
              \vbox{\hrule width5pt height0.4pt depth0pt
                    \kern -2pt
                    \hbox{\kern 2.3pt
                          \vrule width0.4pt height6pt depth0pt
                          }
                    \kern -2pt
                    \hrule width5pt height0.4pt depth0pt}%
                    \kern 1pt
           }
            {
              \kern 1pt%
              \raise 1pt
              \vbox{\hrule width4.3pt height0.4pt depth0pt
                    \kern -1.8pt
                    \hbox{\kern 1.95pt
                          \vrule width0.4pt height5.4pt depth0pt
                          }
                    \kern -1.8pt
                    \hrule width4.3pt height0.4pt depth0pt}%
                    \kern 1pt
            }
            {
              \kern 0.5pt%
              \raise 1pt
              \vbox{\hrule width4.0pt height0.3pt depth0pt
                    \kern -1.9pt  
                    \hbox{\kern 1.85pt
                          \vrule width0.3pt height5.7pt depth0pt
                          }
                    \kern -1.9pt
                    \hrule width4.0pt height0.3pt depth0pt}%
                    \kern 0.5pt
            }
            {
              \kern 0.5pt%
              \raise 1pt
              \vbox{\hrule width3.6pt height0.3pt depth0pt
                    \kern -1.5pt
                    \hbox{\kern 1.65pt
                          \vrule width0.3pt height4.5pt depth0pt
                          }
                    \kern -1.5pt
                    \hrule width3.6pt height0.3pt depth0pt}%
                    \kern 0.5pt
            }
        }}

  \def\mm{{\mathchoice
                       {
                             \kern 1pt
               \raise 1pt    \vbox{\hrule width5pt height0.4pt depth0pt
                                  \kern 2pt
                                  \hrule width5pt height0.4pt depth0pt}
                             \kern 1pt}
                       {
                            \kern 1pt
               \raise 1pt \vbox{\hrule width4.3pt height0.4pt depth0pt
                                  \kern 1.8pt
                                  \hrule width4.3pt height0.4pt depth0pt}
                             \kern 1pt}
                       {
                            \kern 0.5pt
               \raise 1pt
                            \vbox{\hrule width4.0pt height0.3pt depth0pt
                                  \kern 1.9pt
                                  \hrule width4.0pt height0.3pt depth0pt}
                            \kern 1pt}
                       {
                           \kern 0.5pt
             \raise 1pt  \vbox{\hrule width3.6pt height0.3pt depth0pt
                                  \kern 1.5pt
                                  \hrule width3.6pt height0.3pt depth0pt}
                           \kern 0.5pt}
                       }}

\catcode`@=11
\def\un#1{\relax\ifmmode\@@underline#1\else
        $\@@underline{\hbox{#1}}$\relax\fi}
\catcode`@=12

\let\du=\du                     

\def\a{\alpha}
\def\b{\beta}

\def\d{\delta}

\def\f{\phi}

\def\l{\lambda}

\def\q{\theta}

\def\s{\sigma}

\def\ve{\varepsilon}

\def\vq{\vartheta}

\def\cd{{\cal D}}

\def\bo{{\raise-.5ex\hbox{\large$\Box$}}}               
\def\TH{{\raise.2ex\hbox{$\displaystyle \bigodot$}\mskip-4.7mu \llap H \;}}
\def\face{{\raise.2ex\hbox{$\displaystyle \bigodot$}\mskip-2.2mu \llap {$\ddot
        \smile$}}}                                      

\def\Bar#1{\overline{#1}}                       
\def\leftrightarrowfill{$\mathsurround=0pt \mathord\leftarrow \mkern-6mu
        \cleaders\hbox{$\mkern-2mu \mathord- \mkern-2mu$}\hfill
        \mkern-6mu \mathord\rightarrow$}
\def\dvec#1{\vbox{\ialign{##\crcr
        \leftrightarrowfill\crcr\noalign{\kern-1pt\nointerlineskip}
        $\hfil\displaystyle{#1}\hfil$\crcr}}}           
\def\dt#1{{\buildrel {\hbox{\LARGE .}} \over {#1}}}     

\def\frac#1#2{{\textstyle{#1\over\vphantom2\smash{\raise.20ex
        \hbox{$\scriptstyle{#2}$}}}}}                   
\def\sfrac#1#2{{\vphantom1\smash{\lower.5ex\hbox{\small$#1$}}\over
        \vphantom1\smash{\raise.4ex\hbox{\small$#2$}}}} 
\def\bfrac#1#2{{\vphantom1\smash{\lower.5ex\hbox{$#1$}}\over
        \vphantom1\smash{\raise.3ex\hbox{$#2$}}}}       
\def\afrac#1#2{{\vphantom1\smash{\lower.5ex\hbox{$#1$}}\over#2}}    
\def\on#1#2{\mathop{\null#2}\limits^{#1}}               
\def\bvec#1{\on\leftarrow{#1}}                  

\def\[{\lfloor{\hskip 0.35pt}\!\!\!\lceil}
\def\]{\rfloor{\hskip 0.35pt}\!\!\!\rceil}
\def\Lag{{\cal L}}
\def\du#1#2{_{#1}{}^{#2}}

\def\fracm#1#2{\hbox{\large{${\frac{{#1}}{{#2}}}$}}}
\def\ha{{\fracmm12}}

\def\Tr{{\rm Tr}}

\def\un{\underline}
\def\fracmm#1#2{{{#1}\over{#2}}}

\def\low#1{{\raise -3pt\hbox{${\hskip 0.75pt}\!_{#1}$}}}

\def\Dot#1{\buildrel{_{_{\hskip 0.01in}\bullet}}\over{#1}}
\def\dt#1{\Dot{#1}}

\newskip\humongous \humongous=0pt plus 1000pt minus 1000pt
\def\caja{\mathsurround=0pt}
\def\eqalign#1{\,\vcenter{\openup2\jot \caja
        \ialign{\strut \hfil$\displaystyle{##}$&$
        \displaystyle{{}##}$\hfil\crcr#1\crcr}}\,}
\newif\ifdtup

\parskip=\medskipamount                 
\thicklines                         

\begin{document}
\thispagestyle{empty}

{\hbox to\hsize{
\vbox{\noindent June 2004 \hfill hep-th/0405278 }}}

\noindent
\vskip1.3cm
\begin{center}

{\Large\bf Non-anticommutative N=2 Supersymmetric SU(2) Gauge Theory 
~\footnote{Supported in part by the JSPS and the Volkswagen Stiftung}}
\vglue.2in

Sergei V. Ketov~\footnote{Email address: ketov@phys.metro-u.ac.jp}
and Shin Sasaki~\footnote{Email address: shin-s@phys.metro-u.ac.jp}

{\it Department of Physics, Faculty of Science\\
          Tokyo Metropolitan University\\
         1--1 Minami-osawa, Hachioji-shi\\
            Tokyo 192--0397, Japan}
\end{center}
\vglue.2in
\begin{center}
{\Large\bf Abstract}
\end{center}

\noindent
We calculate the component Lagrangian of the four-dimensional
{\it non-anticommutative} (with a singlet deformation parameter) and fully 
N=2 supersymmetric gauge field theory with the simple gauge group $SU(2)$. We 
find that the deformed (classical) scalar potential is unbounded from below, in
contrast to the undeformed case.

\newpage

\section{Introduction}

Supersymmetric gauge field theories in {\it Non-AntiCommutative} (NAC) 
superspace \cite{nac} recently became the area of intensive study, 
see e.g., ref.~\cite{rec}. The motivation is two-fold, at least. Firstly, 
those NAC-deformed field theories naturally arise from superstrings in certain
 supergravity backgrounds and, second, they are natural extensions of the 
ordinary supersymmetric gauge field theories (formulated in the standard 
(anti)commutative superspace).

Here we always assume that merely a chiral part of the fermionic superspace 
coordinates becomes NAC, whereas the other superspace coordinates still 
(anti)commute (in some basis). This is only possible when the anti-chiral 
fermionic coordinates $(\bar{\q})$ are not complex conjugates to the chiral 
ones, $\bar{\q}\neq(\q)^*$, which is the case in Euclidean or Atiyah-Ward 
spacetimes with the signature $(4,0)$ and $(2,2)$, respectively. The Euclidean
signature is relevant to instantons and superstrings \cite{rec}, whereas the 
Atiyah-Ward signature is relevant to the critical N=2 string models \cite{ov} 
or the supersymmetric self-dual gauge field theories \cite{mary}.

Extended supersymmetry offers more opportunities depending upon how much of
supersymmetry one wants to preserve, as well as which NAC deformation (e.g., 
a singlet or a non-singlet) and which operators (the supercovariant derivatives
or the supersymmetry generators) one wants to employ in the Moyal-Weyl star
product \cite{nac,2nac}. The $N=(1,1)$ (or just $N=2$) extended supersymmetry 
is special since it allows one to choose a singlet NAC deformation that 
preserves all the fundamental symmetries \cite{fs,ks1}. Indeed, the most 
general nilpotent deformation of $N=(1,1)=2\times(\ha,\ha)$ supersymmetry is 
given by 
$$ \{ \q_i^{\a},\q_j^{\b}\}_{\star}=\d^{(\a\b)}_{(ij)}C^{(\a\b)}+
2iP\ve^{\a\b}\ve_{ij}\quad {\rm (no~sum!)}~,\eqno(1.1)$$
where $\a,\b=1,2$ are chiral spinor indices, $i,j=1,2$ are the indices of the 
internal R-symmetry group $SU(2)_R$, while $C^{\a\b}$ and $P$ are some 
constants. Taking only a singlet deformation to be non-vanishing, $P\neq 0$, 
and using the chiral supercovariant N=2 superspace derivatives  $D_{i\a}$ in 
the Moyal-Weyl star product,
$$ A\star B = A\exp\left( iP\ve^{\a\b}\ve^{ij}\bvec{D}{}_{i\a}\vec{D}{}_{j\b}
\right)B~~,\eqno(1.2)$$ 
allows one to keep manifest N=2 supersymmetry, Lorentz invariance and 
R-invariance, as well as (undeformed) gauge invariance (after some non-linear
field redefinition) \cite{fs,ks1}. The star product (1.2) matching those
conditions is unique, while it requires $N=2$.

We choose flat Euclidean spacetime for definiteness, but continue to use the 
notation common to N=2 superspace with Minkowski spacetime signature, as it is
 becoming increasingly customary in the current literature 
(see ref.~\cite{book} for details about our notation). Our NAC N=2 superspace 
with the coordinates $(x^m,\q^{i}_{\a},\bar{\q}_{i}^{\dt{\a}})$ is defined by
eq.~(1.1), with $C^{\a\b}=0$ and $P\neq 0$, as the only non-trivial 
(anti)commutator amongst the N=2 superspace coordinates. This choice preserves
most fundamental features of N=2 supersymmetry, such as G-analyticity 
\cite{fs}.

A NAC-deformed (non-abelian) supersymmetric gauge field theory can also be 
rewritten to the usual form, with the standard gauge transformations of field 
components, i.e. as some kind of effective action, after certain (non-linear) 
field redefinition, known as the Seiberg-Witten map ({\it cf.} ref.~\cite{sw}).
In the case of the $P$-deformed N=2 super-Yang-Mills theory such (non-abelian) 
map was calculated in ref.~\cite{fs} with the following result for the 
effective anti-chiral N=2 superfield strength $\Bar{W}$:
$$  \Bar{W}_{\rm NAC}= \fracmm{\Bar{W}}{1+P\Bar{W}}~~,\eqno(1.3)$$
where $\Bar{W}$ is the standard (Lie algebra-valued) N=2 anti-chiral superfield
strength. The effective N=2 superspace action reads
$$ S_{\rm NAC}=-\fracm{1}{2}\int d^4x\low{R} 
d^4\bar{\q}\,\Tr\, \Bar{W}^2_{\rm NAC} \equiv -\fracm{1}{2} \int d^4x\low{R}
d^4\bar{\q}\,\Tr f(\Bar{W})~, \eqno(1.4)$$
whose structure function $f(\Bar{W})$ is thus given by \cite{fs}
$$ f(\Bar{W})= \left(\fracmm{\Bar{W}}{1+P\Bar{W}}\right)^2~~.\eqno(1.5)$$
It is non-trivial to calculate eq.~(1.4) in components because of the need
to perform the (non-abelian) group-theoretical trace (the Lagrangian is no 
longer quadratic in $\Bar{W}$!). In this Letter we consider only the simplest
non-abelian gauge group $SU(2)$. Some partial results in the $SU(3)$ case 
will be reported elsewhere \cite{ks3}. The component action of the $P$-singlet 
NAC-deformed N=2 supersymmetric $U(1)$ gauge field theory is fully 
straightforward to calculate from eqs.~(1.4) and (1.5) --- see 
refs.~\cite{fs,ks1}.

Our paper is organized as follows. In sect.~2 we perform the $SU(2)$ 
group-theoretical trace in eq.~(1.4) and find yet another effective function 
of the colorless variable $\Tr(\Bar{W}{}^2)$  that governs the component 
action. In sect.~3 we give the full component action of the $P$-deformed N=2 
supersymmetric $SU(2)$ gauge field theory. In sect.~4 we focus on the scalar 
potential of the deformed theory. Sect.~5 is our conclusion.

\section{Calculation of the $SU(2)$ trace}

The anti-hermitian $SU(2)$ matrices in the adjoint (vector) representation are 
$$ T^1 = \left( \begin{array}{ccc} 0 & 0 & 0 \\ 0 & 0 & -1 \\ 0 & 1 & 0
\end{array}\right)~~,\quad
T^2 = \left( \begin{array}{ccc} 0 & 0 & 1 \\ 0 & 0 & 0 \\ -1 & 0 & 0
\end{array}\right)~~,\quad
T^3 = \left( \begin{array}{ccc} 0 & -1 & 0 \\ 1 & 0 & 0 \\ 0 & 0 & 0
\end{array}\right)~~.\eqno(2.1)$$
Their matrix elements are given by the $SU(2)$ structure constants, 
$(T^a)^{bc}=-\ve^{abc}$, where $\ve^{abc}$ is the totally antisymmetric
Levi-Civita symbol, $\ve^{123}=1$. The matrices (2.1) obey the $SU(2)$ Lie
algebra, $\[T^a,T^b\]=\ve^{abc}T^c$, where $a,b,\ldots=1,2,3$.

The $SU(2)$ trace in eqs.~(1.4) and (1.5) is given by
$$\eqalign{
\Tr\left( \fracmm{\Bar{W}{}^aT^a}{{\bf 1} + P\Bar{W}{}^bT^b}\right)^2  &~ =
\Tr\left[ \Bar{W}{}^aT^a({\bf 1} + P\Bar{W}{}^bT^b)^{-1}
\Bar{W}{}^cT^c({\bf 1} + P\Bar{W}{}^dT^d)^{-1}\right] \cr
&~ = \Tr\left[ \Bar{W}{}^aT^a \sum^{\infty}_{n=0}(-)^n(P\Bar{W}{}^bT^b)^n
\Bar{W}{}^cT^c \sum_{m=0}^{\infty}(-)^m(P\Bar{W}{}^dT^d)^m\right] \cr
&~ = \Tr \sum^{\infty}_{m,n=0} (-)^{n+m}P^{n+m}(\Bar{W}{}^aT^a)^{n+m+2} \cr
&~ = \sum^{\infty}_{n=0}(-)^n(n+1)P^{n}\Tr(\Bar{W}{}^aT^a)^{n+2}~.\cr}
\eqno(2.2)$$
It is straightforward to calculate $(m\geq 1)$
$$ \Tr(\Bar{W}{}^aT^a)^{2m}=2(-)^m(\Bar{W}{}^a\Bar{W}{}^a)^m\quad {\rm and}
\quad  \Tr(\Bar{W}{}^aT^a)^{2m+1}=0~~.\eqno(2.3)$$
Hence, eq.~(2.2) is equal to
$$ 2\sum^{\infty}_{n=0}(-)^{n+1}(2n+1)P^{2n}(\Bar{W}{}^a\Bar{W}{}^a)^{n+1}=
\fracmm{2}{P^2}\sum^{\infty}_{n=1}(-)^{n}(2n-1)(P^2\Bar{W}{}^a\Bar{W}{}^a)^n$$
$$\eqalign{
{} &  =\fracmm{4}{P^2}\sum^{\infty}_{n=1}n(-P^2\Bar{W}{}^a\Bar{W}{}^a)^n-
 \fracmm{2}{P^2}\sum^{\infty}_{n=1}(-P^2\Bar{W}{}^a\Bar{W}{}^a)^n \cr
{} &  =\fracmm{4}{P^2}\fracmm{(-P^2)\Bar{W}{}^a\Bar{W}{}^a}{(1+
P^2\Bar{W}{}^a\Bar{W}{}^a)^2}+ 
\fracmm{2}{P^2}\fracmm{P^2\Bar{W}{}^a\Bar{W}{}^a}{(1+
P^2\Bar{W}{}^a\Bar{W}{}^a)} \cr
{} & = \fracmm{-2\Bar{W}{}^a\Bar{W}{}^a+2P^2(\Bar{W}{}^a\Bar{W}{}^a)^2}{(1+
P^2\Bar{W}{}^a\Bar{W}{}^a)^2}\equiv g(\Bar{W}{}^2)~~.\cr}\eqno(2.4) $$ 
In the limit $P\to 0$ we obtain the usual (undeformed) $SU(2)$-based $N=2$ 
super-Yang-Mills theory. Having introduced the gauge coupling 
constant $g\low{\rm YM}$ explicitly and rescaled the action by the factor of 
$1/{g^2\low{\rm YM}}$, we can also consider another limit, $P\neq 0$ but 
$g\low{\rm YM}\to 0$, that gives rise to the undeformed (free and abelian) 
N=2 gauge theory.

\section{The Lagrangian in components}  

The standard (undeformed) $N=2$ gauge superfield strength
$\Bar{W}$ is defined by the anticommutator of two gauge- and super-covariant 
spinor derivatives in N=2 superspace,
$$ \{ \cd^i_{\a}, \cd^j_{\b} \}= -2\ve^{ij}\ve_{\a\b}\Bar{W}~,
\eqno(3.1)$$ 
and it obeys the N=2 superfield Bianchi identities,
$$ \cd_{i\a}\Bar{W}=0\qquad {\rm and}\qquad \Bar{\cd}_{ij}\Bar{W}=
\cd_{ij}W~.\eqno(3.2)$$
The non-abelian N=2 superfield $\Bar{W}$ is thus a {\it covariantly} 
anti-chiral N=2 superfield, being not an N=2 anti-chiral one. However, the
 composite `colorless' N=2 superfield 
$\Bar{W}{}^2\equiv\Bar{W}{}^a\Bar{W}{}^a$ is N=2 anti-chiral, 
$D_{i\a}\Bar{W}{}^2=0$, so that it can be expanded with respect to the 
anticommuting N=2 superspace variables $\bar{\q}_{i\dt{\a}}$ 
(in the anti-chiral N=2 basis) as follows ({\it cf.} ref.~\cite{ks1}):
$$\Bar{W}{}^2 = U +V_{\dt{\a}i}\bar{\q}^{\dt{\a}i}+X_{ij}\bar{\q}^{ij}
+Y_{\dt{\a}\dt{\b}}\bar{\q}^{\dt{\a}\dt{\b}}
+Z_{\dt{\a}i}(\bar{\q}^3)^{\dt{\a}i} + L\bar{\q}^4~,\eqno(3.3)$$
where we have introduced its (composite) field components 
$(U,V_{\dt{\a}i},X_{ij},Y_{\dt{\a}\dt{\b}},Z_{\dt{\a}i},L)$.

We define the covariant field components of the N=2 superfield $\Bar{W}$  by 
covariant differentiation of $\Bar{W}$,
$$ \Bar{W}|=\bar{\f}~,\quad \Bar{\cd}_{i\dt{\a}}\Bar{W}|=
\bar{\l}_{i\dt{\a}}~,\quad \Bar{\cd}_{ij} \Bar{W}|=D_{ij}~,\quad
\Bar{\cd}_{\dt{\a}\dt{\b}} \Bar{W}|=F_{\dt{\a}\dt{\b}}~,\eqno(3.4)$$
where $|$ denotes the leading ($\q$- and $\bar{\q}$-independent) component of
an N=2 superfield. In particular, $F_{\dt{\a}\dt{\b}}^a=
(\s^{mn})_{\dt{\a}\dt{\b}}F^a_{mn}$ is the anti-self-dual part of the 
Yang-Mills field strength $F_{mn}^a$, the chiral spinors (gaugino) 
$\bar{\l}^a_{i\dt{\a}}$ transform as a doublet under $SU(2)_R$ and as a triplet
under $SU(2)$, the scalars (higgs) $\bar{\f}^a$ form a triplet under 
$SU(2)$, whereas the $SU(2)_R\times SU(2)$ double triplet 
$D^a_{ij}=D^a_{ji}$ are the auxiliary fields.

The composites of eq.~(3.3) in terms of the field components (3.4) read as 
follows: 
$$U=\bar{\f}^a\bar{\f}^a,~~
V_{\dt{\a}i}=2\bar{\l}^a_{\dt{\a}i}\bar{\f}^a,~~
X_{ij}= 2\left(\bar{\f}^aD^a_{ij}-\bar{\l}^{\dt{\a}a}_i
\bar{\l}^{a}_{j\dt{\a}}\right)~,~~
Y_{\dt{\a}\dt{\b}}  =2\left(\bar{\f}^aF^a_{\dt{\a}\dt{\b}}-
\bar{\l}^{ia}_{\dt{\a}}\bar{\l}^{a}_{i\dt{\b}}\right)~,$$
$$Z_{i\dt{\a}} = 4i\bar{\f}^a(\tilde{\s}^m)\du{\dt{\a}}{\a}
\cd_m\l^a_{i\a} + \bar{\l}^{ja}_{\dt{\a}}D_{ij} 
- \bar{\l}^{\dt{\b}a}_{i}F^a_{\dt{\a}\dt{\b}}~~,\eqno(3.5a)$$
and
$$\eqalign{
 L  = ~&~ -2\bar{\f}^a\cd_m\cd^m\f^a -i\bar{\l}^a_{i\dt{\a}}
(\tilde{\s}^m)^{\dt{\a}\a}\cd_m\l^{ia}_{\a} 
 + \ve^{abc}\l^{ia\a}\bar{\f}^b\l^c_{i\a} +
\ve^{abc}\bar{\l}^{a}_{i\dt{\a}}\f^b\bar{\l}^{\dt{\a}ic} \cr
~&~ + \fracmm{1}{48}D^a_{ij}D^{aij}- \fracmm{1}{12}F^{mna-}F^{a-}_{mn} 
 -\fracm{1}{2}\f^a\bar{\f}^b\f^c\bar{\f}^d\ve^{abf}\ve^{cdf}~,
\cr}\eqno(3.5b)$$
where $\cd_m$ are the usual gauge-covariant derivatives (in the adjoint), 
$F^-_{mn}$ is the anti-self-dual part of $F_{mn}$. The last composite field 
$L$ is nothing but the usual N=2 super-Yang-Mills Lagrangian,  
$L=\Lag_{\rm SYM}$.

The deformed N=2 gauge theory action in undeformed N=2 superspace is
given by eqs.~(1.4) and (1.5),
$$ S=-\fracm{1}{2}\Tr\,\int d^4x\low{R}d^4\bar{\q}\,f(\Bar{W})=-\fracm{1}{2}
\int d^4x\low{R}\bar{D}{}^4\Tr\,f(\Bar{W})=-\fracm{1}{2}
\int d^4x\low{R}\Tr(\Bar{\cd}{}^4f(\Bar{W}))~.\eqno(3.6)$$
Here $\Bar{\cd}{}^4$ is the gauge-covariant extension of $\Bar{D}{}^4$, 
$$ \bar{\cd}{}^4 \equiv \fracmm{1}{4!}\ve^{\un{i}\un{k}\un{m}\un{n}}
\Bar{\cd}_{\un{i}}\Bar{\cd}_{\un{k}}\Bar{\cd}_{\un{m}}\Bar{\cd}_{\un{n}}
= \fracmm{1}{96}\left(\Bar{\cd}_{ij}\Bar{\cd}^{ij}-\Bar{\cd}_{\dt{\a}\dt{\b}}
\Bar{\cd}^{\dt{\a}\dt{\b}}\right)-W{}^2~,\eqno(3.7)$$
where we have also used the composite indices, $\un{i}=(i\dt{\a})$, to 
introduce our definition of $\bar{\cd}{}^4$. It is most straightforward to 
compute the component Lagrangian specified by a (colorless) effective function 
$g(\Bar{W}{}^2)$ when using an identity
$$\eqalign{
\int d^4\bar{\q}\, g(\Bar{W}{}^2) = &~
 g'(\bar{\f}{}^2)L + g''(\bar{\f}{}^2)\left[ -V_{i\dt{\a}}Z^{i\dt{\a}}+
2X_{ij}X^{ij}-2Y_{\dt{\a}\dt{\b}}Y^{\dt{\a}\dt{\b}}\right] \cr
 & + g'''(\bar{\f}{}^2)\left[ -V_{i\dt{\a}} V^i_{\dt{\b}}Y^{\dt{\a}\dt{\b}}
+V_{i\dt{\a}} V_j^{\dt{\a}}X^{ij}\right]+ g''''(\bar{\f}{}^2)V^4~,\cr}
\eqno(3.8)$$
where the primes denote differentiations with respect to $\bar{\f}{}^2\equiv
 \bar{\f}^a\bar{\f}^a$.  We  find now useful to introduce 
more book-keeping notation,
$$\eqalign{
 (\bar{\l}^2)_{ij} & 
=\bar{\l}^a_{i\dt{\a}}\bar{\l}_j^{\dt{\a}a} \quad {\rm and}
\quad (\bar{\l}^2)_{\dt{\a}\dt{\b}}=
\bar{\l}^a_{i\dt{\a}}\bar{\l}^{ia}_{\dt{\b}}~,\cr
(\bar{\l}^2)_{ij}^{ab} & =\bar{\l}^a_{i\dt{\a}}\bar{\l}_j^{\dt{\a}b} 
\quad {\rm and} \quad 
(\bar{\l}^2)^{ab}_{\dt{\a}\dt{\b}}
=\bar{\l}^a_{i\dt{\a}}\bar{\l}^{ib}_{\dt{\b}}~~,\cr
(\bar{\l}^2)^{ab} & = \bar{\l}^a_{i\dt{\a}}\bar{\l}^{i\dt{\a}b}
 \quad {\rm and} \quad 
(\bar{\l}^4)^{abcd}  = \bar{\l}^{1a}_{\dt{1}}\bar{\l}^{1b}_{\dt{2}}  
\bar{\l}^{2c}_{\dt{1}} \bar{\l}^{2d}_{\dt{2}}~,\cr}\eqno(3.9)$$
together with some related identities \cite{book},
$$ \bar{\l}^4\equiv \fracm{1}{12}(\bar{\l}^2)_{ij}(\bar{\l}^2)^{ij}=
-\fracm{1}{12}(\bar{\l}^2)_{\dt{\a}\dt{\b}}(\bar{\l}^2)^{\dt{\a}\dt{\b}}~.
\eqno(3.10a)$$ 
and
$$ (\bar{\l}^4)^{ab}\equiv \fracm{1}{12}(\bar{\l}^2)_{ij}^{ab}
(\bar{\l}^2)^{ij}=-\fracm{1}{12}(\bar{\l}^2)^{ab}_{\dt{\a}\dt{\b}}
(\bar{\l}^2)^{\dt{\a}\dt{\b}}~~.\eqno(3.10b)$$

Equations (2.2), (2.4) and (3.8) imply that the full component Lagrangian 
$\Lag_{\rm deformed}$ of the $P$-deformed N=2 supersymmetric $SU(2)$ gauge  
theory is governed by a single function,
$$ F(\bar{\f}^2)\equiv -\fracm{1}{2}g'(\bar{\f}^2)=
\fracmm{1-3P^2\bar{\f}^2}{(1+P^2\bar{\f}^2)^3}= 1 -6P^2\bar{\f}^2+O(P^4\f^4)~.
\eqno(3.11)$$

Putting everything together gives rise to our main result:
$$\eqalign{ 
\Lag_{\rm deformed~SYM} = ~& F(\bar{\f}^2)\Lag_{\rm SYM}
+2F'(\bar{\f}^2) \left[ -4i\bar{\f}^a\bar{\f}^b
(\bar{\l}^a_i\tilde{\s}^m\cd_m\l^{ib}) + 
\bar{\f}^a(\bar{\l}^2)^{ab}_{ij}D^{ijb} \right. \cr
~& +8\bar{\f}^aD^a_{ij}(\bar{\l}^2)^{ij}
 +4\bar{\f}^a\bar{\f}^bD^a_{ij}D^{bij} 
 -\bar{\f}^a(\bar{\l}^2)^{ab}_{\dt{\a}\dt{\b}}
(\tilde{\s}^{mn})^{\dt{\a}\dt{\b}}F^{b-}_{mn} \cr
~& \left.- 8\bar{\f}^aF^{a-}_{mn}(\tilde{\s}^{mn})_{\dt{\a}\dt{\b}}
(\bar{\l}^2)^{\dt{\a}\dt{\b}}-128\bar{\f}^a\bar{\f}^bF^{a-}_{mn}F^{mnb-}
+ 96 \bar{\l}^4 \right] \cr
~&  +8F''(\bar{\f}^2) \left[ -\bar{\f}^a\bar{\f}^b\bar{\f}^c
(\bar{\l}^2)^{ab}_{\dt{\a}\dt{\b}}(\tilde{\s}^{mn})^{\dt{\a}\dt{\b}}
F^{c-}_{mn}  + \bar{\f}^a\bar{\f}^b\bar{\f}^c(\bar{\l}^2)^{ab}_{ij}D^{ijc}
\right. \cr
~& \left. + 24 \bar{\f}^a\bar{\f}^b(\bar{\l}^4)^{ab}\right] 
 + 16 F'''(\bar{\f}^2) \bar{\f}^a\bar{\f}^b\bar{\f}^c\bar{\f}^d
(\bar{\l}^4)^{abcd}~,\cr}\eqno(3.12)$$  
where the undeformed N=2 Lagrangian $\Lag_{\rm SYM}$ is given by eq.~(3.5b).

\section{Scalar potential}

Perhaps, the most interesting part of the deformed Lagrangian (3.12) is its
scalar potential
$$ V_{\rm deformed} = -\fracmm{g_{\rm YM}^2}{4}F(\bar{\f}^2)
\Tr\[\f,\bar{\f}\]^2\equiv F(\bar{\f}^2)V_{\rm SYM}~,\eqno(4.1)$$
where we have explicitly introduced the gauge coupling constant, and the
undeformed (non-abelian) N=2 super-Yang-Mills scalar potential $V_{\rm SYM}$.
Equations (3.11) and (3.12) now imply
$$\eqalign{
 V_{\rm deformed} &~ =\fracm{1}{2}
g_{\rm YM}^2F(\bar{\f}^2) \ve^{abf}\f^a\bar{\f}^b
\ve^{cdf}\f^c\bar{\f}^d \cr
&~ = \fracmm{g_{\rm YM}^2(1-3P^2\bar{\f}^2)}{2(1+P^2\bar{\f}^2)^3}\left[
 \f^2\bar{\f}^2-(\f^a\bar{\f}^a)^2\right]~.\cr}\eqno(4.2)$$
When using the notation
$$ (\f^a\bar{\f}^a)^2= \f^2\bar{\f}^2\cos^2\vq \eqno(4.3)$$
we easily find
$$  V_{\rm deformed}(\f,\bar{\f})=\fracm{1}{2}
g_{\rm YM}^2\f^2\bar{\f}^2\sin^2\vq\,
\fracmm{1-3P^2\bar{\f}^2}{(1+P^2\bar{\f}^2)^3}~~.\eqno(4.4)$$

The scalar potential $V_{\rm SYM}$ of the undeformed N=2 super-Yang-Mills 
theory is bounded from below (actually, non-negative), while the undeformed 
(and degenerate) classical vacua are given by solutions to the equation
$$ \[\f,\bar{\f}\]=0~.\eqno(4.5)$$

In the deformed case under consideration the fields $\f$ and $\bar{\f}$ are 
real and independent, while the $P$-deformation gives rise to the extra factor 
$F(\bar{\f}^2)$ in eqs.~(4.1) and (4.4). Choosing $P^2<0$ gives rise to a 
singular scalar potential at $\bar{\f}^2=-P^{-2}$. We choose $P^2>0$ to get a 
non-singular scalar potential at finite values of $\f$ and $\bar{\f}$.

\begin{figure}[t]
\begin{center}
\vglue.1in
\makebox{
\epsfxsize=2in
\epsfbox{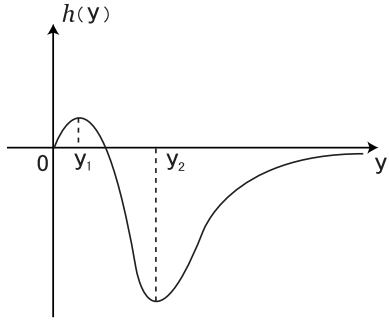}
 }
\caption{\small Graph of the function $h(y)=y(1-3P^2y)(1+P^2y)^{-3}$}
\end{center}
\end{figure}

A graph of the relevant function $h(y)\equiv y(1-3P^2y)(1+P^2y)^{-3}$ with
$y\equiv\bar{\f}^2$ is given in  Fig.~1. The function $h(y)$ is bounded from 
below and from above, as long as $y\geq 0$, with its maximum at 
$y_1=P^{-2}(4-\sqrt{13})/3$ and its minimum at $y_2=P^{-2}(4+\sqrt{13})/3$.  
Nevertheless, the full scalar potential is unbounded from below, just because
the function $h(y)$ can take negative values at some finite $\bar{\f}$ 
(including its value at the minimum), while one can still have 
$\sin^2\vq> 0$ and $\f\to \infty$, which implies  $V\to -\infty$. Therefore, 
the classical $P$-deformed $SU(2)$-based N=2 supersymmetric gauge field theory
 does  not have a stable vacuum.  

\section{Conclusion}

It is worth noticing that the {\it non-abelian} $SU(2)$-based N=2 NAC 
Lagrangian found in this Letter is very {\it different} from the abelian 
$U(1)$-based Lagrangian with the same $P$-deformation and star product 
\cite{fs,ks1}, despite of the fact that both originate from the same 
Seiberg-Witten map (1.3). The non-abelian NAC Lagrangian in components is
 governed by another function -- see eqs.~(2.2) and (2.4).

Our considerations in this paper were entirely classical. It would be 
interesting to investigate the role of quantum corrections, both in quantum
field theory and in string theory (e.g., by using geometrical engineering).
It is particularly intriguing to know whether quantum corrections can
stabilize the classical vacuum.

\end{document}